\documentclass[a4paper,10pt,twoside]{cpc-hepnp}
\usepackage{CJK,upgreek,fancyhdr}
\usepackage{multicol}
\usepackage{graphicx}
\usepackage{booktabs}
\usepackage{amssymb,bm,mathrsfs,bbm,amscd}
\usepackage[tbtags]{amsmath}
\usepackage{lastpage}

\begin{document}
\begin{CJK*}{GB}{gbsn}

\fancyhead[c]{\small Chinese Physics C~~~Vol. xx, No. x (201x) xxxxxx}
\fancyfoot[C]{\small 010201-\thepage}

\footnotetext[0]{Received 20 June 2016}

\title{Photoproduction of hidden-charm states in the $\gamma p \to \bar{D}^{*0} \Lambda^+_c$ reaction near threshold
\thanks{Supported by the Major State Basic Research Development Program in China (No. 2014CB845400), the
National Natural Science Foundation of China (Grants No. 11475227,
No. 11275235, No. 11035006) and the Chinese Academy of Sciences (the
Knowledge Innovation Project under Grant No. KJCX2-EW-N01). It is
also supported by the Youth Innovation Promotion Association CAS
(No. 2016367) and by the Open Project Program of State Key
Laboratory of Theoretical Physics, Institute of Theoretical Physics,
Chinese Academy of Sciences, China (No. Y5KF151CJ1) }}

\author{%
Yin Huang (»ÆÒø)$^{1,2,3;1)}$\email{huangy2014@lzu.cn}%
\quad Ju-Jun Xie (л¾Û¾ü)$^{1,3;2)}$\email{xiejujun@impcas.ac.cn}%
\quad Jun He (ºÎ¾ü)$^{1,3}$ \quad Xurong Chen (³ÂÐñÈÙ)$^{1,3}$
\\ \quad Hong-Fei Zhang (Õźè·É)$^{2,3}$ } \maketitle

\address{%
$^1$ Research Center for Hadron and CSR Physics, Lanzhou University
and Institute of Modern Physics of CAS, Lanzhou
730000,China \\
$^2$ School of Nuclear Science and Technology,
Lanzhou University, Lanzhou 730000, China \\
$^3$ Institute of Modern physics, Chinese Academy of Sciences,
Lanzhou 730000, China }

\begin{abstract}

We report on a theoretical study of the hidden charm $N^*_{c
\bar{c}}$ states in the $\gamma p \to \bar{D}^{*0} \Lambda^+_c$
reaction near threshold within an effective Lagrangian approach. In
addition to the contributions from the $s$-channel nucleon pole, the
$t$-channel $D^0$ exchange, the $u$-channel $\Lambda^+_c$ exchange
and the contact term, we study the contributions from the $N^*_{c
\bar{c}}$ states with spin-parity $J^P = 1/2^-$ and $3/2^-$. The
total and differential cross sections of the $\gamma p \to
\bar{D}^{*0} \Lambda^+_c$ reaction are predicted. It is found that
the contributions of these $N^*_{c \bar{c}}$ states give clear peak
structures in the total cross sections. Thus, this reaction is
another new platform to study the hidden-charm states. It is
expected that our model calculation may be tested by future
experiments.

\end{abstract}



\footnotetext[0]{\hspace*{-3mm}\raisebox{0.3ex}{$\scriptstyle\copyright$}2013
Chinese Physical Society and the Institute of High Energy Physics
of the Chinese Academy of Sciences and the Institute
of Modern Physics of the Chinese Academy of Sciences and IOP Publishing Ltd}%

\begin{multicols}{2}

\section{Introduction}

The study of hadron states beyond the traditional quark model,
namely the exotic states, has been a hot topic in hadron physics,
and these studies will improve our understanding of non-perturbative
QCD. With the experimental progress on this issue over the past
decade, more and more evidence in the heavy quark sector indicates
possible candidates for exotic states, which are called the $XYZ$
states. However, experimental evidence for exotic pentaquark
baryon states has been missing for a long time. Recently, the LHCb
Collaboration reported two hidden-charm pentaquark states
$P_c^{+}(4380)$ and $P_c^{+}(4450)$ in the $J/\psi p$ invariant mass
spectrum from the $\Lambda_b^{0} \to J/\psi p K^{-}$
decay~\cite{Aaij:2015tga}, and their masses and widths are
$M_{P_c(4380)} = 4380 \pm 8 \pm 29$ MeV, $\Gamma_{P_c(4380)} = 205
\pm 18 \pm 86$ MeV, $M_{P_c(4450)} = 4449.8 \pm 1.7 \pm 2.5$ MeV,
$\Gamma_{P_c(4450)} = 39 \pm 5 \pm 19$ MeV. Since the two states
were reported from the final state $J/\psi p$ invariant mass
distribution, the isospin of the $P_c(4380)$ and $P_c(4450)$ is
$1/2$ and they are ideal candidates for pentaquark states with
quark content of $c\bar{c}uud$. According to partial wave
analysis, the preferred spin-parity $J^P$ of the $P_c(4380)$ and the
$P_c(4450)$ are either $\frac{3}{2}^{-}$ and $\frac{5}{2}^{+}$ or
$\frac{3}{2}^{+}$ and $\frac{5}{2}^{-}$, respectively. After this,
various explanations for the $P_c(4380)$ and $P_c(4450)$ states were
proposed from the theoretical side (see more details in a recent
review paper~\cite{Chen:2016qju}). However, the structure of these
two states is still an open question, and none of the explanations in
the literature have been accepted unanimously.

Actually, predictions of hidden charm baryon states have been
made before.  In Ref.~\cite{Yang:2011wz}, loosely bound hidden
charm molecular baryons composed of anti-charmed meson and charmed
baryon were obtained with the one-boson-exchange model. In
Refs.~\cite{Wu:2010jy,Wu:2010vk}, the interaction between various
anti-charmed mesons and charmed baryons plus decay channels in the
light sector was studied within the framework of the coupled-channel
unitary approach. Several dynamically generated $N^{*}_{c\bar{c}}$
and $\Lambda^{*}_{c\bar{c}}$ resonances with hidden charm were
predicted with mass above 4 GeV and width smaller than 100
MeV~\cite{Wu:2010jy,Wu:2010vk}. Using Heavy Quark Spin Symmetry and
the local hidden gauge approach, further studies were done in
Ref.~\cite{Xiao:2013yca} and similar results to those of
Refs.~\cite{Wu:2010jy,Wu:2010vk} were found in the study of the
interaction of the $J/\psi N$, $\bar{D}^* \Lambda_c$, $\bar{D}^*
\Sigma_c$, $\bar{D} \Sigma^*_c$, and $\bar{D}^* \Sigma^*_c$ coupled
channels. In the $I = 1/2$ sector, three states with $J^P = 1/2^-$
and three states with $J^P = 3/2^-$ were dynamically
generated~\cite{Xiao:2013yca} (see table I of that paper). Based on
the results of Refs.~\cite{Wu:2010jy,Wu:2010vk,Xiao:2013yca}, the
molecule nature, $\bar{D}^* \Sigma_c$-$\bar{D}^* \Sigma^*_c$
molecular states, of the above two $P_c$ states was proposed in
Refs.~\cite{Roca:2015dva,Roca:2016tdh}. Besides, the nature of the
$P_c(4380)$ was investigated in Ref.~\cite{Shen:2016tzq}, and it was
found that the decays of the $P_c(4380)$ to $\bar{D}^{*}\Lambda_c$
and $J/\psi p$ are very different for $P_c(4380)$ being the
$\bar{D}\Sigma^*_c$ and the $\bar{D}^*\Sigma_c$ molecule states.
Hence, the study of the $P_c(4380)$ states in the $\bar{D}^*
\Lambda_c$ is important to disentangle the nature of the $P_c(4380)$
states.

There have also been phenomenology studies on the production of
those hidden charm states from scattering processes. In
Refs.~\cite{Huang:2013mua,Wang:2015jsa,Kubarovsky:2015aaa}, the role
of the $P_c$ states were studied in the $\gamma p \to P^+_c \to
J/\psi p$ reaction, while in
Refs.~\cite{Wang:2015xwa,Lu:2015fva,Garzon:2015zva}, the role played
by the hidden charm states were discussed in the $\pi$ beam induced
reactions of $\pi^- p \to \eta_cn$, $\pi^- p \to J/\psi n$, and
$\pi^- p \to D^- \Sigma^+_c$. Along this line, in this work, we
study the role of these hidden charm $N^{*}_{c\bar{c}}$ states in
the $\gamma p \to \bar{D}^{*0} \Lambda_c^{+}$ reaction near the
reaction threshold within the effective Lagrangian approach.
Unfortunately, the couplings of the two observed $P_c(4380)$ and
$P_c(4450)$ states to the $\gamma N$ and $\bar{D}^* \Lambda_c$
channels are unknown, so we will take the work of
Ref.~\cite{Xiao:2013yca} as a reference where the couplings of these
$N^*_{c\bar{c}}$ to $J/\psi p$ and $\bar{D}^* \Lambda_c$ were
obtained. We then provide the total and differential cross sections
of the $\gamma p \to \bar{D}^{*0} \Lambda^+_c$ reaction. It is found
that the differential cross sections for the $N^*_{c\bar{c}}$ states
with different quantum numbers are different, which can be used to
distinguish the quantum numbers of these hidden charm states.

Although
the effective Lagrangian method is a convenient tool to catch the
qualitative features of the $\gamma p \to \bar{D}^{*0} \Lambda^+_c$
reaction, the free parameters in the model give it some
uncontrollable uncertainties. In the present work, based on
phenomenological Lagrangians, we only consider the selected
tree-diagram contributions. However, our calculation offers some
important clues for the mechanisms of the $\gamma p \to \bar{D}^{*0}
\Lambda^+_c$ reaction and makes a first effort to study the role of
the predicted $N^*_{c\bar{c}}$ states in the $\gamma p \to
\bar{D}^{*0} \Lambda^+_c$ reaction. It is expected that future
experimental measurements could test our model and give more
information about the $\gamma p \to \bar{D}^{*0} \Lambda^+_c$
reaction.

This paper is organized as follows. In Section 2, the formalisms and
ingredients for our calculations are listed. The results of total
and differential cross sections and discussions are presented in
Section 3. Finally, a short summary is given in the last section.

\section{Formalisms and ingredients}

\subsection{Feynman diagrams and interaction Lagrangian densities}

We study the $\gamma p \to \bar{D}^{*0} \Lambda_c^{+}$ reaction
within the effective Lagrangian approach, which has been widely
employed to investigate photoproduction processes. The basic
tree level Feynman diagrams for $\gamma p \to \bar{D}^{*0}
\Lambda_c^{+}$ reaction are depicted in Fig.~\ref{feydiagrams},
where the contributions from the hidden charm $N_{\bar{c}c}^{*}$
states [Fig.~\ref{feydiagrams}(a)], the nucleon pole
[Fig.~\ref{feydiagrams}(b)], the $D^{0}$ meson exchange
[Fig.~\ref{feydiagrams}(c)], the $\Lambda_c^{+}$ exchange
[Fig.~\ref{feydiagrams}(d)], and the contact term
[Fig.~\ref{feydiagrams}(e)] are taken into account.

\end{multicols}
\begin{center}
\includegraphics[scale=0.7]{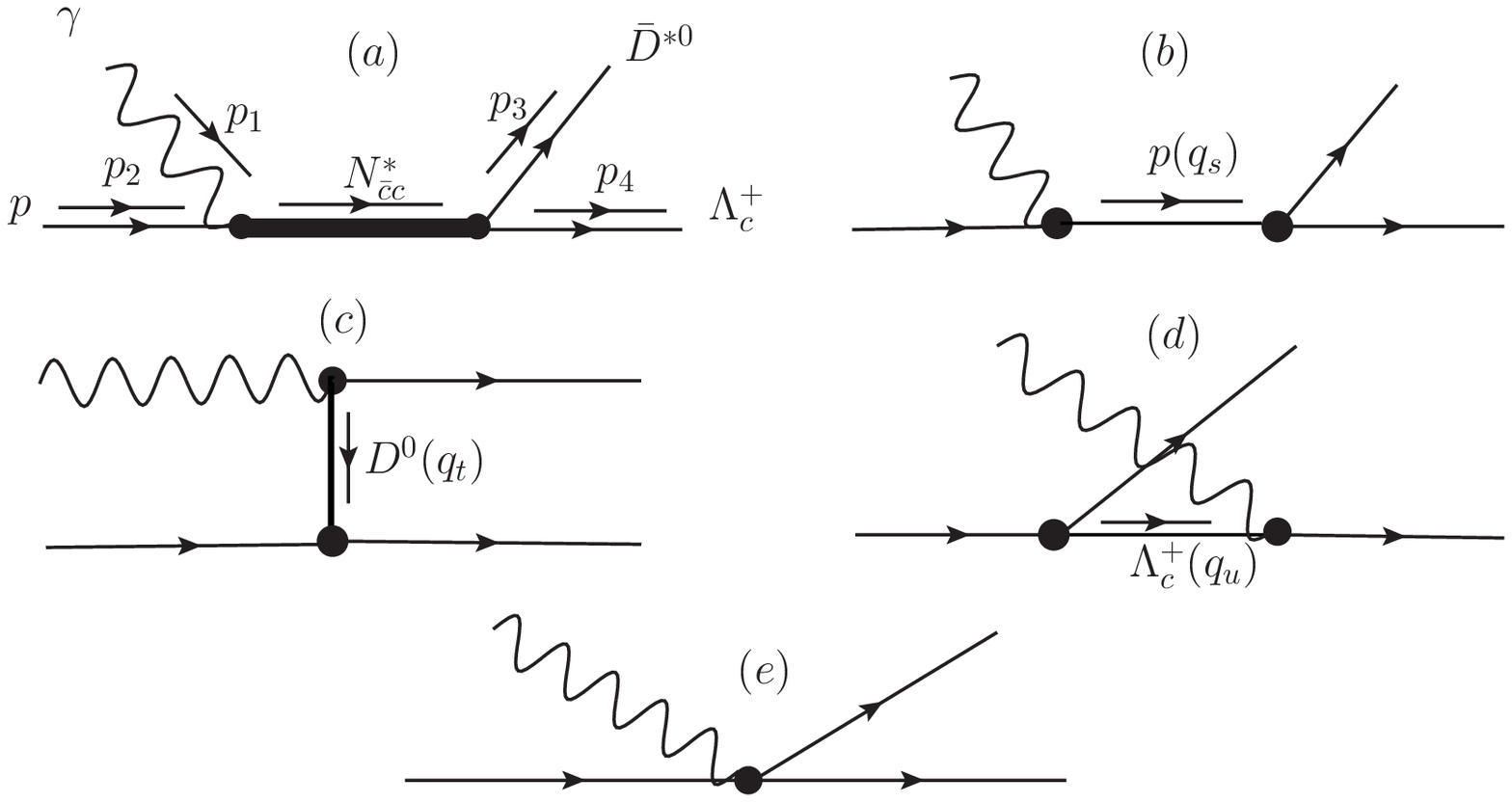}
\figcaption{Feynman diagrams for the $\gamma p \to \bar{D}^{*0}
\Lambda_c^{+}$ reaction.  The contributions from the $s$-channel
$N_{\bar{c}c}^{*}$ resonance (a), $s$-channel nucleon pole (b),
$t$-channel $D^{0}$ exchange (c), $u$-channel $\Lambda_c^{+}$
exchange (d), and contact term (e) are considered. In the first
diagram, we also show the definition of the kinematical ($p_1$,
$p_2$, $p_3$, $p_4$) that we use in the present calculation. In
addition, we use $q_s = p_1 + p_2$, $q_t = p_1 - p_3$, and $q_u =
p_4 - p_1$. \label{feydiagrams}}
\end{center}
\begin{multicols}{2}

To compute the amplitudes of the diagrams shown in
Fig.~\ref{feydiagrams}, the effective Lagrangian densities for the
relevant interaction vertexes are needed. We adopt the effective
Lagrangians as used in
Refs.~\cite{Oh:2007jd,Zou:2002yy,Dong:2014ksa,Dong:2010xv},
\begin{eqnarray}
{\cal{L}}_{\gamma{}N N^{*}_{\bar{c}c}}^{1/2^-} &=& \frac{eh}{2M_N}\bar{N} \sigma_{\mu\nu}\partial^{\nu}A^{\mu}N_{\bar{c}c}^{*}+H.c.,\label{Eq1}\\
{\cal{L}}_{\gamma{}N N^{*}_{\bar{c}c}}^{3/2^-} &=& -ie[\frac{h_1}{2M_N}\bar{N}\gamma_{\nu} + \frac{h_2}{(2M_N)^2}\partial_{\nu}\bar{N} ] \times \nonumber \\
                                                 &&
                                                 {}F^{\mu\nu}N_{\bar{c}c\mu}^{*}+H.c.,\label{Eq2}\\
{\cal{L}}_{N_{\bar{c}c}^{*}N J/\psi}^{1/2^{-}} &=& g^{1/2^{-}}_{N_{\bar{c}c}^{*}NJ/\psi}\bar{N}\gamma_5\gamma_{\mu}N^{*}_{\bar{c}c} \psi^{\mu} + H.c.,\label{Eq3}\\
{\cal{L}}_{{N^{*}_{\bar{c}c}\bar{D}^{*}\Lambda_{c}}}^{1/2^{-}} &=&
g_{{N^{*}_{\bar{c}c}\bar{D}^{*}\Lambda_{c}}}^{1/2^{-}}
\bar{\Lambda}_{c} \gamma_5
\gamma_{\mu}N^{*}_{\bar{c}c}\bar{D}^{*0\mu} + H.c.,\label{Eq4}\\
{\cal{L}}_{N_{\bar{c}c}^{*} N J/\psi}^{3/2^{-}} &=& g_{N_{\bar{c}c}^{*}NJ/\psi}^{3/2^{-}}\bar{N}N^{{*}}_{\bar{c}c\mu} \psi^{\mu} + H.c.,\label{Eq5}\\
{\cal{L}}^{3/2^{-}}_{{N^{*}_{\bar{c}c}\bar{D}^{*}\Lambda_{c}}} &=&
g^{3/2^{-}}_{{N^{*}_{\bar{c}c}\bar{D}^{*0}\Lambda_{c}}}\bar{\Lambda}_cN_{\bar{c}c\mu}^{*}\bar{D}^{*0\mu}
+ H.c. ,\label{Eq6}\\
{\cal{L}}_{\gamma pp} &=& -e\bar{p}(\gamma^{\mu}A_{\mu} -
\frac{k_p}{2M_p}\sigma^{\mu\nu}\partial_{\nu}A_{\mu})p ,\label{Eq8}\\
{\cal{L}}_{\Lambda_c N D}  &=& ig_{\Lambda_c N D}\bar{\Lambda}_c
\gamma_5 N D + H.c.,\label{Eq9}\\
{\cal{L}}_{\Lambda_c N \bar{D}^{*}}  &=& g_{\Lambda_c N \bar{D}^{*}}\bar{\Lambda}_c \gamma^{\mu} N \bar{D}^{*}_{\mu} + H.c.,\label{Eq10}\\
{\cal{L}}_{\gamma \Lambda_c \Lambda_c}  &=&
-e\bar{\Lambda}_c(\gamma^{\mu} A_{\mu} -
\frac{k_{\Lambda_c}}{2M_{\Lambda_c}}\sigma^{\mu\nu}
\partial_{\nu}A_{\mu})\Lambda_c ,\label{Eq11}\\
{\cal{L}}_{D^{*} D \gamma}  &=& \frac{e}{4}g_{D^{*} D \gamma}
\epsilon^{\mu\nu\alpha\beta} F_{\mu\nu} {\cal
D}^{*}_{\alpha\beta}D+H.c.,\label{Eq12}
\end{eqnarray}
where $N^{*}_{\bar{c}c}$ and $N^{*}_{\bar{c}c \mu}$ are the hidden
charm nuclear resonance field with spin-parity $J^{P}=1/2^{-}$ and
$J^{p}=3/2^{-}$, respectively. $N$, $A_{\mu}$, $\Lambda_c$,
$\bar{D}^{*}$, $\psi^{\mu}$ are the nucleon field, photon field,
$\Lambda_c$ field, $D^{*}$ field, and $J/\psi$ field, respectively.
$M_{N}$ and $M_{\Lambda_c}$ are the masses of the nucleon and
$\Lambda^{+}_c(2286)$, while $\epsilon^{\mu\nu\alpha\beta}$ is the
Levi-Civit$\grave{a}$ tensor with $\epsilon^{0123} = 1$. In the
above Lagrangian densities, the definitions of $\sigma_{\mu \nu}$,
$F_{\mu \nu}$, and ${\cal D}^{*}_{\alpha\beta}$ are:
\begin{eqnarray}
\sigma_{\mu\nu} &=&
\frac{i}{2}(\gamma_{\mu}\gamma_{\nu}-\gamma_{\nu}\gamma_{\mu}), \\
F_{\mu\nu} &=& \partial_{\mu} A_{\nu} - \partial_{\nu} A_{\mu}, \\
{\cal D}^{*}_{\alpha\beta} &=&
\partial_{\alpha} D^{*}_{\beta} - \partial_{\beta} D^{*}_{\alpha} .
\end{eqnarray}

\subsection{Coupling constants and form factors}

First, we take the anomalous magnetic momentum $k_p = 1.79$ and
$k_{\Lambda_c} = 0.35$ as used in
Refs.~\cite{Jena:1986xs,Xie:2010yk}. The coupling constants for the
$\Lambda_c p D$ and $\Lambda_c p D^*$ vertexes are taken to be
$g_{\Lambda_c p D} = -13.98$ and $g_{\Lambda_c p \bar{D}^{*}} =
-5.20$ as obtained in
Refs.~\cite{Dong:2014ksa,Dong:2010xv,Xie:2015zga} from $SU(4)$
flavor symmetry.

Second, the coupling constant $g_{D^{*} D \gamma}$ is determined by
the radiative decay width of $D^{*0} \to D^0 \gamma$,
\begin{align}
\Gamma_{D^{*0} \to D^{0} \gamma} &= \frac{e^2 g^2_{D^{*} D
\gamma}}{96 \pi}  M_{D^{*0}}^3 (1 -
\frac{M_{D^0}^2}{M_{D^{*0}}^2})^3. \label{eq:GammaDstarDgamma}
\end{align}
Unfortunately, information about the decay width $\Gamma_{D^{*0} \to
D^{0} \gamma}$ is scarce~\cite{Agashe:2014kda}. Thus, it is necessary
to rely on theoretical predictions, such as those of
Ref.~\cite{Dong:2008gb}, where $\Gamma_{D^{*0} \to D^{0} \gamma} =
26$ keV was deduced from the data on strong and radiative decays of
$D^*$ mesons. From Eq.~\eqref{eq:GammaDstarDgamma} we can easily
obtain $g_{D^{*} D \gamma} = 2.0$ GeV$^{-1}$ with $M_{D^{*0}} =
2.007$ GeV, $M_{D^{0}} =1.865$ GeV and $\Gamma_{D^{*0} \to
D^{0}\gamma} = 26$ keV.

Next, we comment on the coupling constants $h$, $h_1$ and $h_2$ for
$N^*_{c\bar{c}} N \gamma$ vertexes, where their values will be
obtained from the strong coupling constants
$g^{1/2^-}_{N^{*}_{\bar{c}c}NJ/\psi}$ and
$g^{3/2^-}_{N^{*}_{\bar{c}c}NJ/\psi}$ using the vector meson
dominance (VMD) model. We adopt the VMD leading order coupling
between $J/\psi$ and photon:
\begin{equation}
{\cal{L}}_{J/\psi\gamma}=-\frac{eM_{J/\psi}^2}{f_{J/\psi}}
\psi_{\mu}A^{\mu},
\end{equation}
where $M_{J/\psi}$ and $f_{J/\psi}$ denote the mass and the decay
constant of the vector meson $J/\psi$. With the decay width
$\Gamma_{J/\psi \to e^{+}e^{-}} = 5.55$ keV~\cite{Agashe:2014kda},
one obtains the parameter $e/f_{J/\psi}=0.027$.

For the $N^*_{c\bar{c}}$ with $J^P = 3/2^-$, there are two different
coupling structures for the $N^*_{c\bar{c}} N \gamma$ vertex, and
information about the $N^*_{c\bar{c}} \to N \gamma$ transition is
unknown. Thus, it is necessary to rely on previous theoretical
works. As argued in Ref.~\cite{Wang:2015jsa}, for $N^{*}_{\bar{c}c}$
decays into $J/\psi p$, the momentum of the final states are fairly
small compared with the nucleon mass. Thus, the higher partial wave
terms proportional to $(p/M_N)^2$ can be neglected. Nevertheless, in
this work, we will only consider the leading order $s$-wave
$N^*_{c\bar{c}} N \gamma$ coupling and leave the higher partial
waves to further studies. Then, we can relate $h_1$ to the
$g_{N^*_{c\bar{c}} J/\psi N}$ for $N^{*}_{\bar{c}c}$ with $J=3/2^-$.
In the framework of the VMD, the coupling constants $h$ and $h_1$
are related to the strong coupling constant
$g^{1/2^-}_{N^{*}_{\bar{c}c}NJ/\psi}$ and
$g^{3/2^-}_{N^{*}_{\bar{c}c}NJ/\psi}$ as,
\end{multicols}
\begin{eqnarray}
eh &=& g^{1/2^{-}}_{N_{\bar{c}c}^{*}NJ/\psi} \frac{e}{f_{J/\psi}}
\frac{2M_N}{(M_{N^{*}_{\bar{c}c}}^2-M_N^2)M_{J/\psi}}
\sqrt{M_{J/\psi}^2(M_N^2+4M_NM_{N^{*}_{\bar{c}c}}+M_{N^{*}_{\bar{c}c}}^2)+(M_{N^{*}_{\bar{c}c}}^2-M_N^2)^2},   \\
eh_1 &=&
g^{3/2^{-}}_{N_{\bar{c}c}^{*}NJ/\psi}\frac{e}{f_{J/\psi}}\frac{2M_N(M_N+M_{N^{*}_{\bar{c}c}})}{(M_{N^{*}_{\bar{c}c}}^2-M_N^2)M_{J/\psi}}
\sqrt{\frac{6M_{J/\psi}^2M_{N^{*}_{\bar{c}c}}^2+M_N^4-2M_N^2M_{N^{*}_{\bar{c}c}}^2+M_{N^{*}_{\bar{c}c}}^4}{3M_{N^{*}_{\bar{c}c}}^2+M_N^2}}
.
\end{eqnarray}
\begin{multicols}{2}

With the $N^*_{c\bar{c}}$ masses and the values of
$g^{1/2^-}_{N^{*}_{\bar{c}c}NJ/\psi}$ that were obtained in
Ref.~\cite{Xiao:2013yca}, we can easily get the values of $eh$ and
$eh_1$ as shown in Table~\ref{Tab:ccs}, where we show also the
coupling constants $g_{N^*_{c\bar{c}}\bar{D}^* \Lambda_c}$ that we
need to calculate the contributions of the hidden charm
$N^*_{c\bar{c}}$ states in the $\gamma p \to \bar{D}^{*0}
\Lambda^+_c$ reaction.

\end{multicols}
\begin{center}
\tabcaption{Values of the hidden charm $N^*_{c\bar{c}}$ parameters
required for the estimation of the $\gamma p \to \bar{D}^{*0}
\Lambda^+_c$ reaction. Their masses, widths and strong
coupling were predicted in Ref.~\cite{Xiao:2013yca}. The coupling
$eh$ is for $N^*_{c\bar{c}}$ with $J^P = 1/2^-$, while $eh_1$ is for
$N^*_{c\bar{c}}$ with $J^P = 3/2^-$. \label{Tab:ccs}}
\begin{tabular*}{170mm}{@{\extracolsep{\fill}}ccccccc}
\hline \hline $N^*_{c\bar{c}}$    & $J^P$ & Mass (MeV) & Width (MeV)
&  $g_{N^*_{c\bar{c}}\bar{D}^* \Lambda_c}$
&  $g_{N^{*}_{\bar{c}c}NJ/\psi}$    & $eh$ (or $eh_1$) \\
\hline
 $N^*_1$   & $\frac{1}{2}^-$ &  $4262$       & $35.7$    & $0.50$   & $0.76$   & $0.018$     \\
 $N^*_2$   & $\frac{1}{2}^-$ &  $4410$       & $58.9$    & $0.20$   & $1.44$   & $0.034$     \\
 $N^*_3$   & $\frac{1}{2}^-$ &  $4481$       & $57.8$    & $0.12$   & $0.72$   & $0.017$     \\
\hline
 $N^*_4$   & $\frac{3}{2}^-$ &  $4334$       & $38.8$    & $0.28$   & $1.32$   & $0.031$      \\
 $N^*_5$   & $\frac{3}{2}^-$ &  $4417$       & $8.22$    & $0.11$   & $0.53$   & $0.012$      \\
 $N^*_6$   & $\frac{3}{2}^-$ &  $4481$       & $35.8$    & $0.20$   & $1.05$   & $0.024$     \\
\hline \hline
\end{tabular*}
\end{center}
\begin{multicols}{2}

In evaluating the scattering amplitudes of the $\gamma p \to
\bar{D}^{*0} \Lambda^+_c$ reaction, we need to include the form
factors because hadrons are not pointlike particles. For the
$t$-channel $D^{0}$ meson exchange, we adopt here a common scheme
used in many previous
works~\cite{Xie:2015zga,Gao:2010hy,Kim:2011rm},
\begin{align}
{\cal{F}}_{D^0}(q^2_t) = (\frac{\Lambda^2_{D^0} -
M_{D^{0}}^2}{\Lambda^2_{D^0} - q^2_t})^2,
\end{align}
with cutoff parameter $\Lambda_{D^0} = 2.5$ GeV.

For the $s$-channel and $u$-channel processes, we adopt a form
factor~\cite{Gao:2010hy,Kim:2011rm}
\begin{align}
{\cal{F}}_B(q^2_{ex},M_{ex})=\frac{\Lambda_B^4}{\Lambda_B^4+(q_{ex}^2-M_{ex}^2)^2},
\end{align}
where $q_{ex}$ and $M_{ex}$ are the four-momentum and the mass of
the exchanged hadron, respectively. For simplicity, we take
$\Lambda_B=0.5$ GeV ~\cite{Kim:2011rm} for the $s$-channel nucleon
pole, the $u$-channel $\Lambda_c^{+}$ processes and the $s$-channel
resonance exchanges. The numerical
results are not sensitive to the value of $\Lambda_B$ because of the
narrow width of the hidden charm resonances, but the results of the
$t$-channel $D^0$ exchange are sensitive to the value of the cutoff
parameter $\Lambda_{D^0}$. We will come to this point below.

The propagator for the exchanged $D^{0}$ meson used in our
calculation is
\begin{align}
G_{D^0}(q_t) = \frac{i}{q_t^2-M_{D^0}^{2}}.
\end{align}
For the propagator of the spin-$1/2$ and $3/2$ baryon, we take
\begin{align}
G_{1/2}(q)=\frac{i(q\!\!\!/+M)}{q^2-M^2+iM\Gamma},\\
G^{\mu\nu}_{3/2}(q)=\frac{i(q\!\!\!/+M)P^{\mu\nu}(q)}{q^2-M^2+iM\Gamma},
\end{align}
with
\begin{align}
P^{\mu\nu} =&
-g^{\mu\nu}+\frac{1}{3}\gamma^{\mu}\gamma^{\nu}+\frac{1}{3M}(\gamma^{\mu}q^{\nu}-\gamma^{\nu}q^{\mu})
\nonumber
\\& +\frac{2}{3M^2}q^{\mu}q^{\nu},
\end{align}
where $q$ and $M$ stand for the four-momentum and the mass of the
intermediate nucleon pole, $\Lambda_c^{+}$ state and
$N^{*}_{\bar{c}c}$ resonance that are shown in Table~\ref{Tab:ccs},
respectively. Since $q^2<0$ for $u$-channel $\Lambda_c^{+}$
exchange, we take $\Gamma=0$ for $\Lambda_c^{+}$ and also for the
nucleon pole, while for the hidden charm $N^{*}_{\bar{c}c}$
resonance, we take their widths as shown in Table~\ref{Tab:ccs}.

\subsection{Scattering amplitudes}

With the above effective Lagrangian densities, the scattering
amplitudes for the $\gamma{}p\to{}\bar{D}^{*0}\Lambda_c^{+}$
reaction can be obtained straightforwardly. First, we write the
scattering amplitudes for the $N^{*}_{\bar{c}c}$ resonance with
$J^p=1/2^{-}$ and $3/2^{-}$
\begin{align}
{\cal{M}}_a^{1/2}&=ig_{{N^{*}_{\bar{c}c}\bar{D}^{*0}\Lambda_{C}}}^{1/2^{-}}\bar{u}_{\Lambda_c}(p_4,s_4)\gamma_5\gamma_{\mu}\frac{(q\!\!\!/_s+M_{N^{*}})}{s-M_{N^{*}}^2+iM_{N^{*}}\Gamma}\nonumber\\
           &\times{}{\cal{F}}_{N^{*}}(s,M_{N^{*}})\frac{eh_1}{4M_N}\gamma_5(\gamma_{\nu}p\!\!\!/_1-p\!\!\!/_1\gamma_{\nu})\nonumber\\
           &\times{}u(p_2,s_2)\epsilon^{\nu}(p_1,s_1)\epsilon^{*\mu}(p_3,s_3),\\
{\cal{M}}_a^{3/2}&=-ig^{3/2^{-}}_{{N^{*}_{\bar{c}c}\bar{D}^{*0}\Lambda_{C}}}\bar{u}_{\Lambda_c}(p_4,s_4)\frac{(q\!\!\!/_s+M_{N^{*}})P^{\mu\eta}}{s-M_{N^{*}}^2+iM_{N^{*}}\Gamma}\nonumber\\
         &\times{\cal{F}}_{N^{*}}(s,M_{N^{*}})\frac{eh_1}{2M_N}[p_{1\eta}\gamma_{\nu}-p\!\!\!/_{1}g_{\eta\nu}]\nonumber\\
         &\times{}u(p_2,s_2)\epsilon^{*\mu}(p_3,s_3)\epsilon_{\nu}(p_1,s_1),
\end{align}
for Fig.~\ref{feydiagrams}(a), and
\begin{align}
{\cal{M}}_{b}&=-ieg_{\Lambda_cp\bar{D}^{0*}}\frac{1}{s-M_N^2}{\cal{F}}_{N}(s,M_{N})\epsilon^{\nu}(p_1,s_1)\nonumber\\
               &\times \epsilon^{*\mu}(p_3,s_3) \bar{u}_{\Lambda_c}(p_4,s_4)\gamma_{\mu}(q\!\!\!/_s+M_N)\nonumber\\
               &\times[\gamma_{\nu}-\frac{k_p}{4M_N}(\gamma_{\nu}p\!\!\!/_1-p\!\!\!/_1\gamma_{\nu})]u(p_2,s_2),\\
{\cal{M}}_{c}&=-\frac{e}{4}g_{\bar{D}^{0*}D^{0}\gamma}g_{\Lambda_cpD^{0}}\frac{1}{t-M_{D^{0}}^{2}}{\cal{F}}^2_M(t)\nonumber\\
             &\times\epsilon^{\nu}(p_1,s_1)\epsilon^{*\mu}(p_3,s_3)\epsilon^{\eta\sigma\alpha\beta}[p_{1\eta}g_{\sigma\nu}-p_{1\sigma}g_{\eta\nu}]\nonumber\\
             &\times[p_{3\alpha}g_{\mu\beta}-p_{3\beta}g_{\alpha\mu}]\bar{u}_{\Lambda_c}(p_4,s_4)\gamma_5u(p_2,s_2), \\
{\cal{M}}_{d}&=-ieg_{\Lambda_cp\bar{D}^{*}}\frac{1}{u-M_{\Lambda_c}^2}{\cal{F}}_{\Lambda_c}(u,M_{\Lambda_c})\epsilon^{\nu}(p_1,s_1)\nonumber\\
               &\times \epsilon^{*\mu}(p_3,s_3) \bar{u}_{\Lambda_c}(p_4,s_4)[\gamma_{\nu}-\frac{k_{\Lambda_c}}{4M_{\Lambda_c}}(\gamma_{\nu}p\!\!\!/_1-p\!\!\!/_1\gamma_{\nu})]\nonumber\\
               &\times(q\!\!\!/_u+M_{\Lambda_c})\gamma_{\mu}u(p_2,s_2),
\end{align}
for Figs.~\ref{feydiagrams}(b),~\ref{feydiagrams}(c),
and~\ref{feydiagrams}(d), respectively. In the above equations,
$s=q_s^2$, $t=q_t^2$, and $u=q_u^2$ indicate the Mandelstam
variables.

The contact term illustrated in Fig.~\ref{feydiagrams}(e) serves to
keep the full amplitude gauge invariant. For the present
calculation, we adopt the following
form~\cite{Haberzettl:1997jg,Oh:2007jd},
\begin{align}
{\cal{M}}_{e} & =
ieg_{\Lambda_cp\bar{D}^{*}}\bar{u}_{\Lambda_c}(p_4,s_4)\gamma^{\mu}C^{\nu}u(p_2,s_2)
\nonumber \\
& \times \epsilon^{\nu}(p_1,s_1)\epsilon^{*\mu}(p_3,s_3),
\end{align}
where $C^{\nu}$ is expressed as
\begin{align}
C^{\nu}&=(2p_2-p_1)^{\nu}\frac{{\cal{F}}_{N}(s,M_{N})-1}{s-m_{N}^2}{\cal{F}}_{\Lambda_c}(u,M_{\Lambda_c})\nonumber\\
       &+(2p_4-p_1)^{\nu}\frac{{\cal{F}}_{\Lambda_c}(u,M_{\Lambda_c})-1}{u-m_{\Lambda_c}^2}{\cal{F}}_{N}(s,M_{N}).
\end{align}

The differential cross section in the center of mass (c.m.) frame
for the $\gamma p \to \bar{D}^{*0}\Lambda_c^{+}$ reaction is
calculated using the following equation:
\begin{eqnarray}
\frac{d\sigma}{d\cos\theta} = \frac{M_{N}M_{\Lambda_c}}{32 \pi
s}\frac{|\vec{p}^{\rm ~c.m.}_3|}{|\vec{p}^{\rm ~c.m.}_1|}
\sum_{s_1,s_2,s_3,s_4}|{\cal{M}}_{\gamma{}p\to{}\bar{D}^{*0}\Lambda_c^{+}}|^2
,
\end{eqnarray}
where ${\cal{M}}_{\gamma{}p\to{}\bar{D}^{*0}\Lambda_c^{+}} =
{\cal{M}}_a+{\cal{M}}_b+{\cal{M}}_c+{\cal{M}}_d+{\cal{M}}_e$ is the
total scattering amplitude of the $\gamma p \to
\bar{D}^{*0}\Lambda_c^{+}$ reaction, and $\theta$ is the scattering
angle of the outgoing $\bar{D}^{*0}$ meson relative to the beam
direction, while $\vec{p}_1^{\rm ~c.m.}$ and $\vec{p}_3^{\rm ~c.m.}$
are the photon and $\bar{D}^{*0}$ three momenta in the c.m. frame,
which are
\begin{align}
|\vec{p}_1^{\rm ~c.m.}|&=\frac{\lambda^{1/2}(s,0,M_N^2)}{2\sqrt{s}}, \\
|\vec{p}_3^{\rm
~c.m.}|&=\frac{\lambda^{1/2}(s,M_{\bar{D}^{*0}}^2,M_{\Lambda_c}^2)}{2\sqrt{s}},
\end{align}
where $\lambda$ is the K$\ddot{a}$llen function with
$\lambda(x,y,z)=(x-y-z)^2-4yz$.

\section{Numerical results}

\end{multicols}
\begin{center}
\includegraphics[scale=0.31]{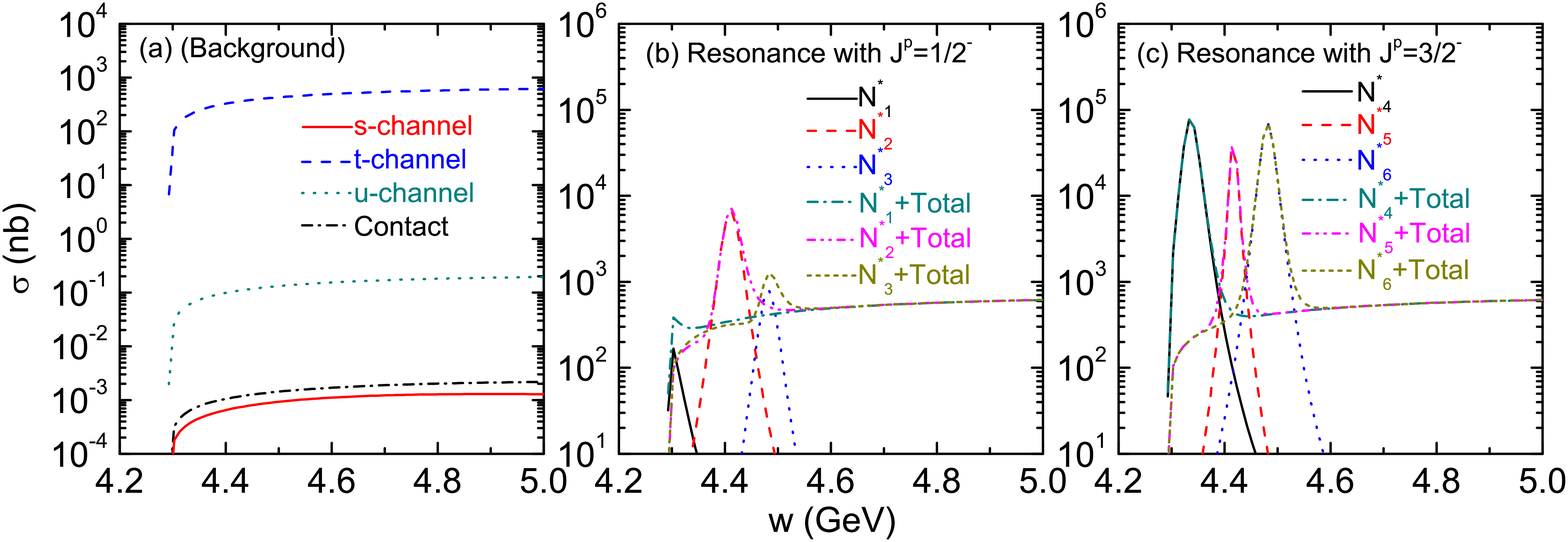}
\figcaption{(Color online) The total cross sections of the
$\gamma{}p\to{}\bar{D}^{*0}\Lambda_c^{+}$ reaction. (a) shows
the cross section of the background contributions including the
$s$-channel nucleon pole (red solid), the $t$-channel $D^{0}$
exchange (blue dashed), $\Lambda_c^{+}(2286)$ as an intermediate state
in the $u$-channel (dark cyan dotted) and the contact term (black dash-dotted). (b) and (c) indicate the contributions of the total and
$s$-channel processes with an $N_{\bar{c}c}^{*}$ resonance of $J=1/2$
and $J=3/2$, respectively. \label{Fig:tcs}}
\end{center}
\begin{multicols}{2}

In this section, we show our theoretical results for the total and
differential cross sections of the
$\gamma{}p\to{}\bar{D}^{*0}\Lambda_c^{+}$ reaction near the
threshold. In Fig.~\ref{Fig:tcs} we show our numerical results for
the total cross section as a function of the invariant mass $W =
\sqrt{s}$ of the $\gamma p$ system. We take the contributions from the
nucleon pole in the $s$-channel, $\Lambda_c^{+}$ exchange in the
$u$-channel, $D^0$ exchange in the $t$-channel, and the contact term as the
background contribution. From Fig.~\ref{Fig:tcs} (a), we see that
the contribution of $D^0$ exchange is larger than the other
background contributions. From Figs.~\ref{Fig:tcs} (b) and
(c), one can see that on top of these background contributions the
peaks of the hidden charm $N_{\bar{c}c}^{*}$ resonances are clearly
seen. In particular, the contributions from $N_{\bar{c}c}^{*}$
resonances with $J^P = 3/2^-$ are significant because of their
narrow widths.

\begin{center}
\includegraphics[scale=0.31]{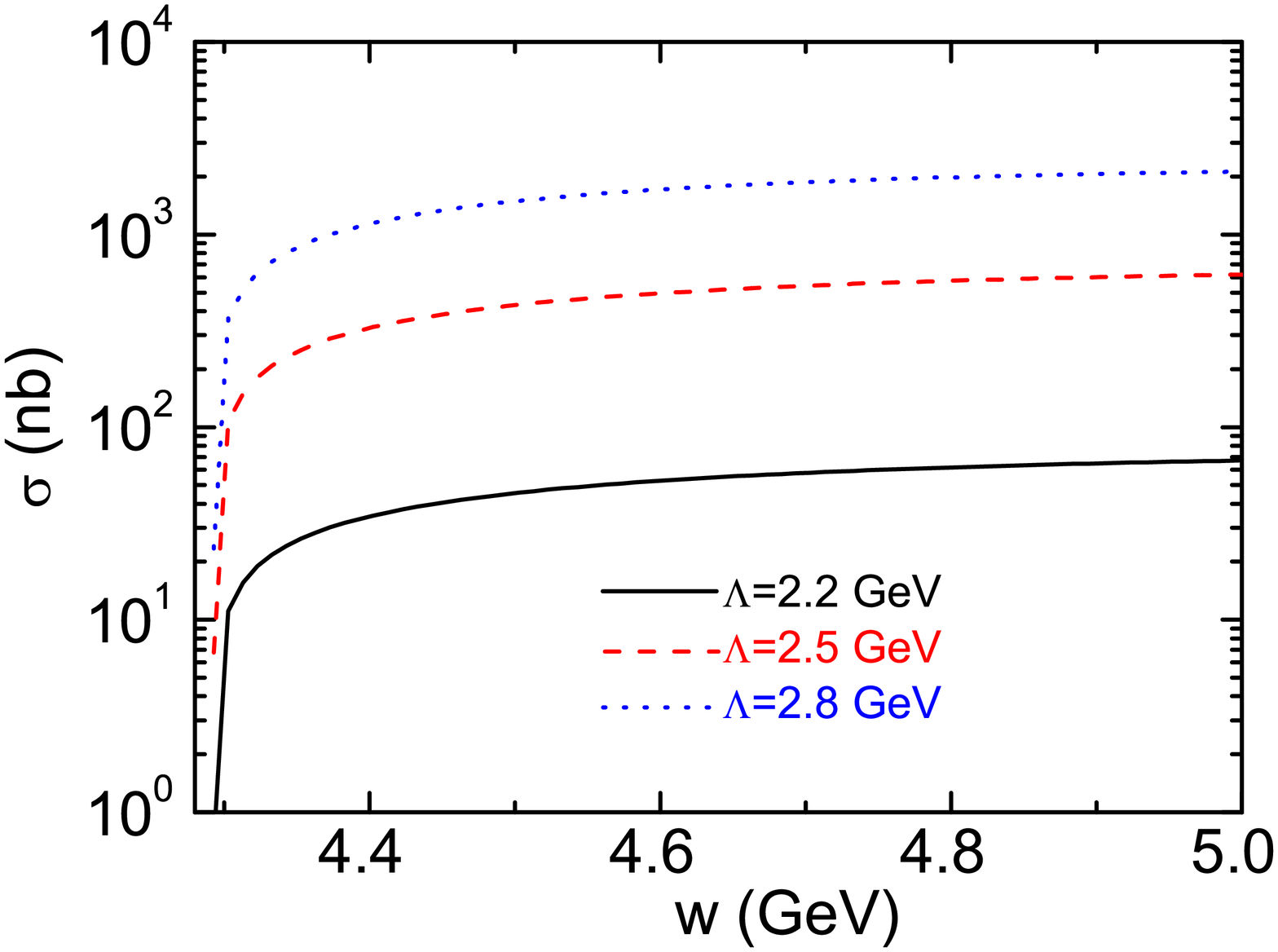}
\figcaption{(Color online) The cross section of the $t$-channel
$D^0$ exchange for the $\gamma{}p\to{}\bar{D}^{*0}\Lambda_c^{+}$
reaction with different $\Lambda_{D^0}$ values. \label{Fig:tcutoff}}
\end{center}

The contribution of $t$-channel $D^0$
exchange is sensitive to the value of the cutoff parameter
$\Lambda_{D^0}$. To see how much it depends on the cutoff parameter
$\Lambda_{D^0}$, in Fig.~\ref{Fig:tcutoff} we show the predicted
cross sections of the $D^0$ exchange with three typical cutoff
parameters $\Lambda_{D^0} = 2.2$, $2.5$, $2.8$ GeV, respectively.
From the figure, it can be clearly seen that the predicted cross
sections have a strong dependence on the cutoff parameter
$\Lambda_{D^0}$. The total cross section reduces by a factor of 10
when $\Lambda_{D^0}$ decreases from $2.5$ to $2.2$ GeV.

In addition to the total cross section, we present in
Figs.~\ref{Fig:dcs1half} and \ref{Fig:dcs3half} the angular
distributions of the $\gamma p \to \bar{D}^{*0} \Lambda^+_c$
reaction at different energies. i.e. $W = 4.3$ GeV, $4.4$ GeV, and
$4.5$ GeV. The results of the $N_{c\bar{c}}$ resonances are shown
with the red-dashed curves. The contributions of backgrounds are
shown with black-solid curves. From Figs.~\ref{Fig:dcs1half} and
\ref{Fig:dcs3half} one can see that the $s$-channel hidden charmed
$N^*_{c\bar{c}}$ contributions are restricted to a narrow kinematic
region because of the narrow widths of these $N^*_{c\bar{c}}$
states. Away from the $N^*_{c\bar{c}}$ resonance region the
$t$-channel $D^0$ exchange plays a dominant role. The contributions
from the $N^*_{c\bar{c}}$ states make the differential cross section
flat because these $N^*_{c\bar{c}}$ states decay into $J/\psi p$ in
$s$-wave, which is different from the diffractive behavior of the
background contributions as shown by Figs.~\ref{Fig:dcs1half}, and
\ref{Fig:dcs3half}. It should be pointed out, if any other possible
states also couple to $\bar{D}^* \Lambda^+_c$ strongly, they will
also cause nondiffractive effects at off-forward angles which can be
measured directly. We hope that this feature may be used to study
the $N_{\bar{c}c}^{*}$ resonance in future experiments.

\end{multicols}
\begin{center}
\includegraphics[scale=0.45]{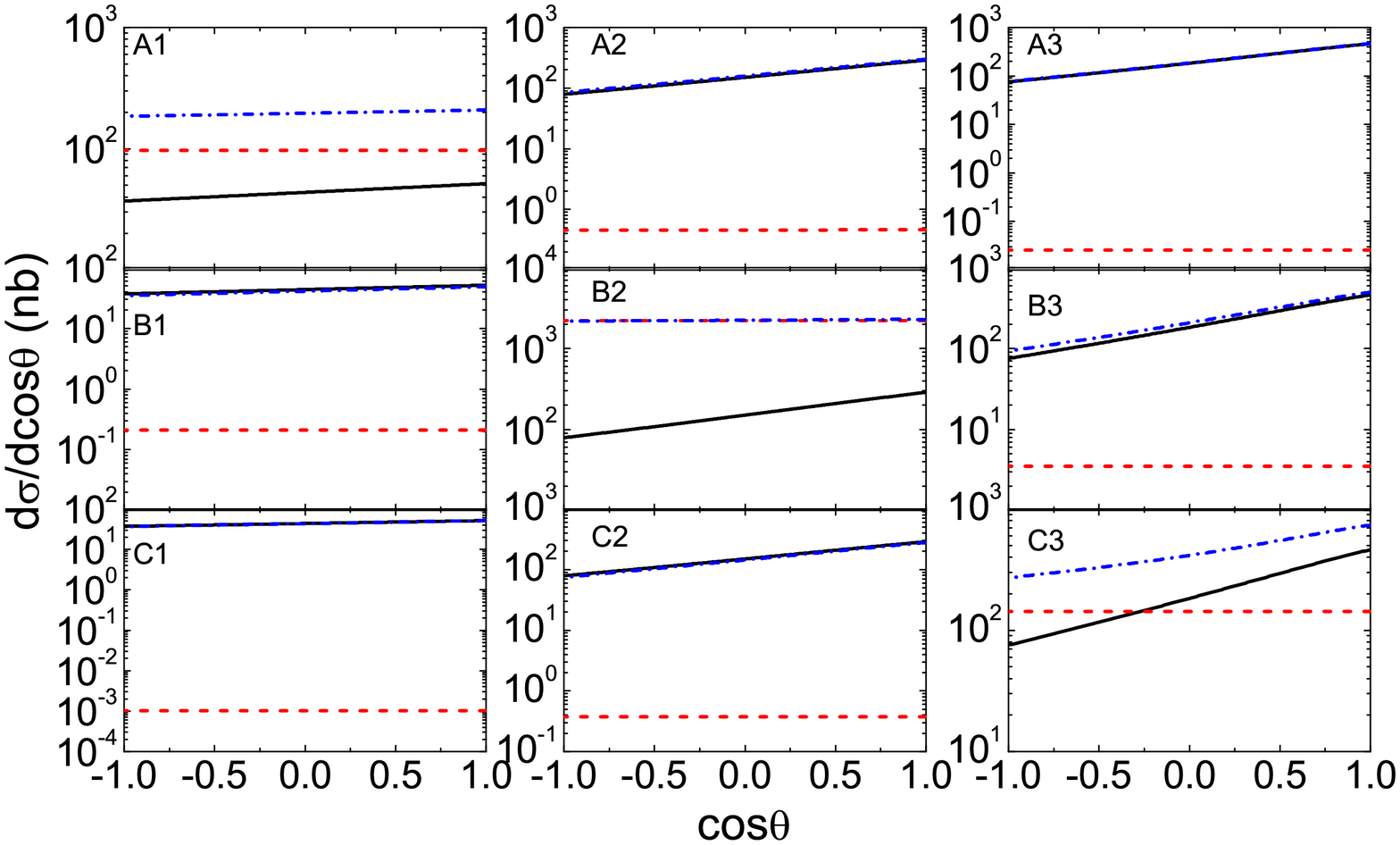}
\figcaption{(Color online) Angular distributions of the $\gamma p
\to \bar{D}^{*0} \Lambda_c^{+}$ reaction including the contributions
of $N^*_{c\bar{c}}$ resonances with $J^P = 1/2^-$ at different
energies: $W = 4.3$ GeV ($A1$, $B1$, and $C1$), $4.4$ GeV ($A2$,
$B2$, and $C2$), and $4.5$ GeV ($A3$, $B3$, and $C3$). $A1$, $A2$,
and $A3$ are for $N_1^{*}$; $B1$, $B2$, and $B3$ are for $N_2^{*}$;
$C1$, $C2$, and $C3$ are for $N_3^{*}$. \label{Fig:dcs1half}}
\end{center}
\begin{multicols}{2}

\end{multicols}
\begin{center}
\includegraphics[scale=0.45]{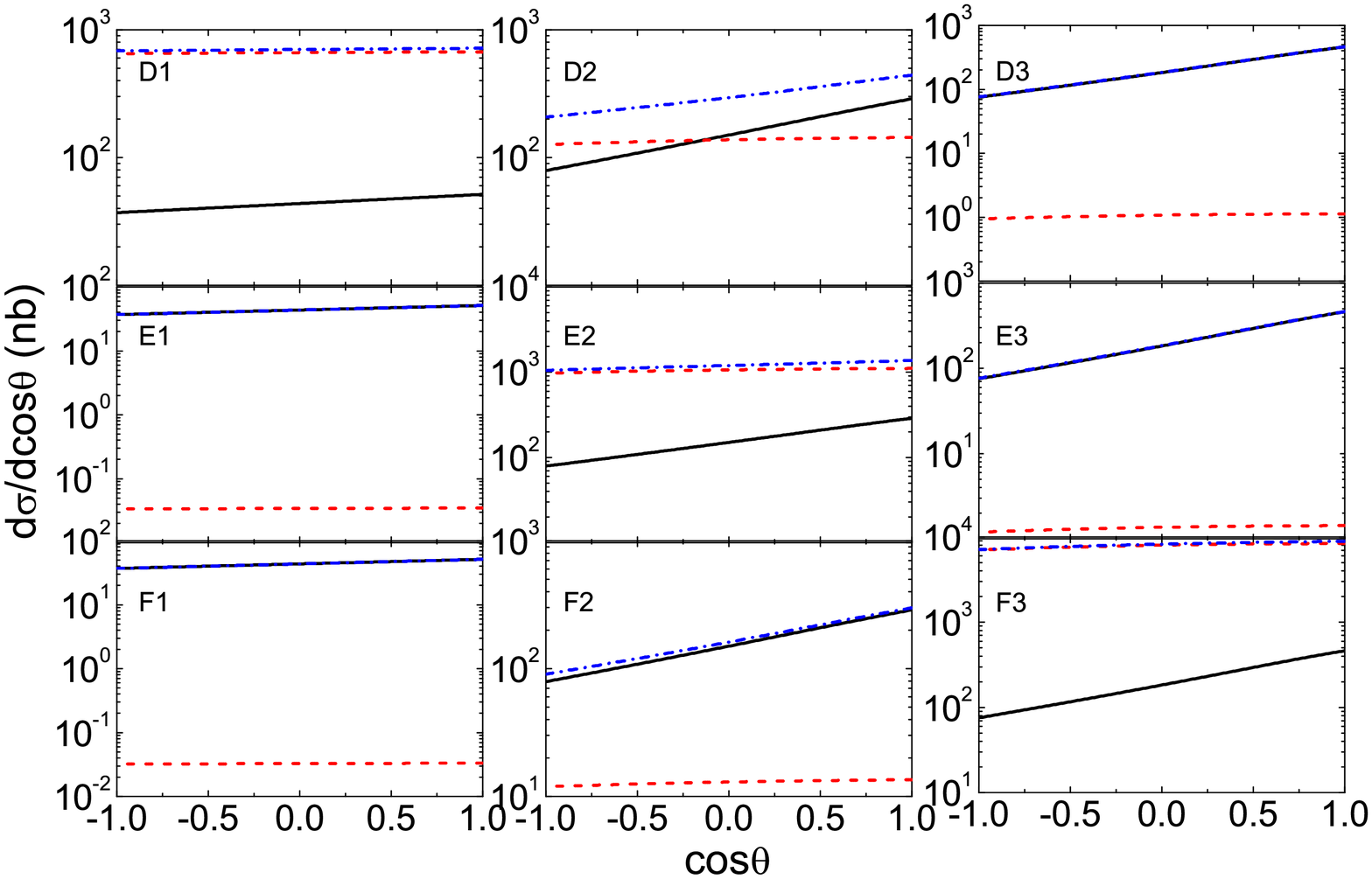}
\figcaption{(Color online) Angular distributions of the $\gamma p
\to \bar{D}^{*0} \Lambda_c^{+}$ reaction including the contributions
of $N^*_{c\bar{c}}$ resonances with $J^P = 3/2^-$ at different
energies: $W = 4.3$ GeV ($D1$, $E1$, and $F1$), $4.4$ GeV ($D2$,
$E2$, and $F2$), and $4.5$ GeV ($D3$, $E3$, and $F3$). $D1$, $D2$,
and $D3$ are for $N_4^{*}$; $E1$, $E2$, and $E3$ are for $N_5^{*}$;
$F1$, $F2$, and $F3$ are for $N_6^{*}$. \label{Fig:dcs3half}}
\end{center}
\begin{multicols}{2}

\section{Summary}

We have studied the total and differential cross sections of the
$\gamma p \to \bar{D}^{*0} \Lambda_c^{+}$ reaction within the effective
Lagrangian model. In addition to the background contributions from
the $s$-channel nucleon pole, $t$-channel $D^{0}$ exchange,
$\Lambda_c^{+}(2286)$ as an intermediate state in the $u$-channel
and the contact term, we checked also the contributions from the
$s$-channel hidden charm $N_{\bar{c}c}^{*}$ resonances which were
dynamically generated from the interaction between charmed mesons
and charmed baryons as reported in
Refs.~\cite{Wu:2010jy,Wu:2010vk,Xiao:2013yca}. The results of the
total cross section show clear peak structures due to the
excitations of the hidden-charm $N^*_{c\bar{c}}$ states. We also evaluated
 the differential cross sections for different energies and
found that the contributions of the $N^*_{c\bar{c}}$ resonances in the
$s$-channel are much different from the background contributions. A
detailed scan of the total and differential cross section in the low
energy region of the $\gamma p \to D^{*0} \Lambda^+_c$ reaction will
help us to study the hidden charm $N^*_{c \bar{c}}$ states.

Furthermore, the two observed $P_c$ states were investigated in
Refs.~\cite{Guo:2015umn,Liu:2015fea,Mikhasenko:2015vca} and it was
shown that the $J/\psi p$ invariant mass spectrum in the $\Lambda_b
\to J/\psi K^- p$ decay can be reproduced by triangle
rescattering due to the reaction dynamics and the peculiar
kinematics. The mechanism is called the triangle singularity, and may
produce threshold enhancements to mimic the behavior of genuine
states or to contribute on top of the genuine states. If these
hidden charm states are genuine states, they should be created in the
$\gamma p \to \bar{D}^{0*} \Lambda^+_c$ reaction, while if they are
the triangle singularity enhancement, they will not appear in the
$\gamma p \to \bar{D}^{0*} \Lambda^+_c$ reaction because the
triangle singularity condition cannot be satisfied here. Therefore,
study of the $\gamma p \to \bar{D}^{0*} \Lambda^+_c$ reaction could
help us to understand the nature of these hidden charm states and
also the two pentaquark $P_c$ states that were observed by
LHCb collaboration~\cite{Aaij:2015tga}.

Finally, we would like to stress that, thanks to the important role
played by the $s$-channel resonant contribution in the $\gamma p \to
\bar{D}^{*0} \Lambda^+_c$ reaction, future experimental data for
this reaction can be used to improve our knowledge of hidden charm
$N^*_{c \bar{c}}$ properties, which are at present poorly known.
This work constitutes a first step in this direction.

\end{multicols}
\vspace{-1mm} \centerline{\rule{80mm}{0.1pt}} \vspace{2mm}

\begin{multicols}{2}

\end{multicols}

\clearpage
\end{CJK*}
\end{document}